\documentclass[mathleft
]{an}
\usepackage{graphicx}
\usepackage{times}
\begin{document}

\Pagespan{789}{}
\Yearpublication{2006}%
\Yearsubmission{2005}%
\Month{11}%
\Volume{999}%
\Issue{88}%

\title{The Innermost Extremes of Black Hole Accretion}

\author{A.C. Fabian\inst{1}\fnmsep\thanks{Corresponding author:
  \email{acf@ast.cam.aac.uk}\newline}
}
\titlerunning{Extremes of BH accretion}
\authorrunning{A.C. Fabian}
\institute{
Institute of Astronomy,
University of Cambridge,
Madingley Road,
Cambridge CB3 0HA,
United Kingdom}

\received{2015}
\accepted{2015}
\publonline{later}

\keywords{Black Holes, active galaxies, X-rays}

\abstract{The inner 20 gravitational radii around the black hole at
  the centre of luminous Active Galactic Nuclei and stellar mass Black
  Hole Binaries are now being routinely mapped by X-ray
  spectral-timing techniques. Spectral blurring and reverberation of
  the reflection spectrum are key tools in this work. In the most
  extreme AGN cases with high black hole spin, when the source appears
  in a low state, observations probe the region within 1 gravitational
  radius of the event horizon. The location, size and operation of the
  corona, which generates the power-law X-ray continuum, is also being
  revealed. }

\maketitle

\section{Introduction}

The extremes of Black Hole accretion concentrate on the innermost
regions of the accretion flow where the bulk of the luminosity is
released. Here I briefly review the basic picture of this region that
has emerged from X-ray observations of luminous accreting black holes.

The accretion flow is dominated by an accretion disc (Fig.~1) which
for luminous objects is optically thick and physically thin, provided
they are below the Eddington limit. Quasi-thermal blackbody emission
from the disc accounts for much of the UV emission in Active Galactic
Nuclei (AGN) and the softer X-ray emission from stellar-mass Black
Hole Binaries (BHB). A variable power-law continuum is also always
seen in AGN and in most spectral states of BHB. This is attributed to
a spatially compact region called the corona, named after the Solar
Corona, although the coronae in BHB and AGN are billions to many
trillions of times more luminous. The corona is presumably powered by
magnetic fields extending from the disc (Galeev, Rosner \& Vaiana
1979). Spectral-timing studies discussed below indicate that, in many
well-studied luminous objects, the accretion disc extends down to the
innermost stable circular orbit (ISCO) and the corona lies above its
central region (Reynolds 2014).

\begin{figure}
\includegraphics[width=0.95\columnwidth]{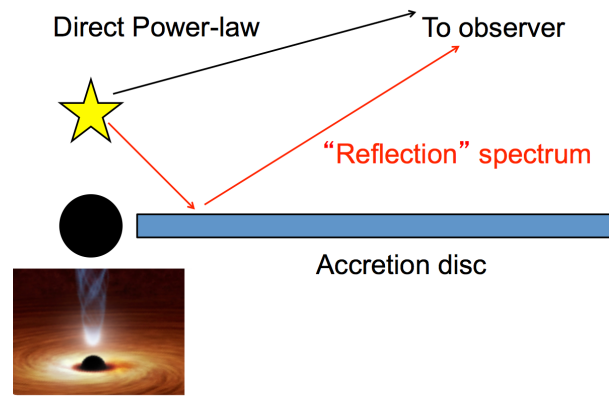}
\caption{Schematic representation of a thin accretion disc orbiting a
  black hole with the corona located on the rotation axis. The
  coronal power-law continuum irradiates the accretion disc producing
  the reflection spectrum. The extra light path in the reflection
  component produces a lag between variations in the direct power-law
  and reflection. The inset shows an artistic representation of the
  setup showing a possible collimated outflow or jet.   
}
\label{label1}
\end{figure}

Irradiation of the accretion disc by the power-law X-ray continuum
leads to fluorescent and backscattered emission from the disc which
produces another spectral component, known as the reflection spectrum
(Fabian \& Ross 2010; Garcia et al 2011,13). This is rich in emission
lines, particularly Fe K, which appear blurred and broadened by
Doppler shifts and gravitational redshift (Fabian et al 1989; Laor
1991; Fig.~2). Measurements of the blurring extent of the reflection
spectrum enable the geometry, particularly radii, and the flow pattern
of the disc and corona to be mapped in gravitational units
(e.g. gravitational radii, $r_{\rm g}=GM/c^2$).

Changes in the power, location or particle outflow in the corona lead
to delayed variations or reverberation in the reflection spectrum
(Fabian et al 1989; Stella 1990; Reynolds et al 1999). The lag time is
just the extra light crossing time from the corona to disc and
out. Measurement of time lags turns the length scales, particularly
the coronal height, into physical units (Uttley et al 2014).

The region that is being mapped often lies within the innermost
$20r_{\rm g}$ around the black hole and sometimes the innermost
$2r_{\rm g}$ when the black hole is spinning rapidly and the coronal
height is small. Note that 50\% of the accretion 
power from  a rapidly spinning black hole emerges
from the inner $5r_{\rm g}$ (Thorne 1994). X-ray observations of
luminous, rapidly spinning, black hole sources probe closest to the
event horizon. The effects of strong gravity, such as strong
gravitational redshift, strong light bending, strong Shapiro effect
and dragging of inertial frames, are all involved in modelling the
spectral-timing results.

\section{The Reflection Spectrum}

The reflection spectrum typically consists of 3 parts: a soft excess,
broad iron line and Compton hump. It is produced by the competition
between electron scattering and photoelectric absorption of the
incident photon. Above about 20 keV, Compton scattering dominates
creating the Compton hump, and below that energy photoelectric
absorption dominates so the spectra drops away (Lightman \& White
1988). When the surface is ionized, electron scattering is again
important together with re-emission through emission lines, causing
the soft excess which rises into the FUV to conserve energy (Ross \&
Fabian 1993, 2005; Garcia et al 2011, 2013; Fig.~2). The intrinsic reflection
spectrum, which would be observed by an observer corotating with part
of the disk, has many emission lines below 7~keV. Relativistic
blurring redshifts and broadens the spectrum (Fabian et al 1989) so
that it appears to be much smoother, particularly if the black hole
spin is high, when blurring is most pronounced (Laor et al 1991;
Dauser et al 2013).

\begin{figure}
\includegraphics[width=0.95\columnwidth]{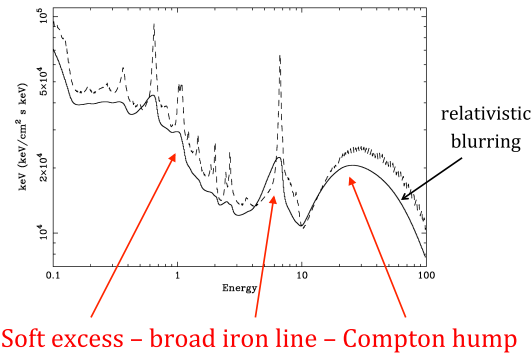}
\caption{The reflection spectrum showing the intrinsic emission
  (dashed) and the relativistically blurred emission  (solid line).
}
\label{label1}
\end{figure}

The whole reflection spectrum is now commonly seen in many
observations of AGN and BHB (e.g. Risaliti et al 2013; Marinucci et al
2014; Miller et al 2013a). The advent of the focussing telescope on NuSTAR
(Harrison et al 2012) means that the Compton hump is now clearly seen
and good models of the whole spectrum can be routinely made and used
to determine parameters such as spin. Even AGN with strong variable
absorption, such as NGC1365, can be modelled and shown to have an
underlying reflection spectrum similar to that found in unobscured
sources (Risaliti et al 2013; Walton et al 2014).

\begin{figure}
\includegraphics[width=0.95\columnwidth]{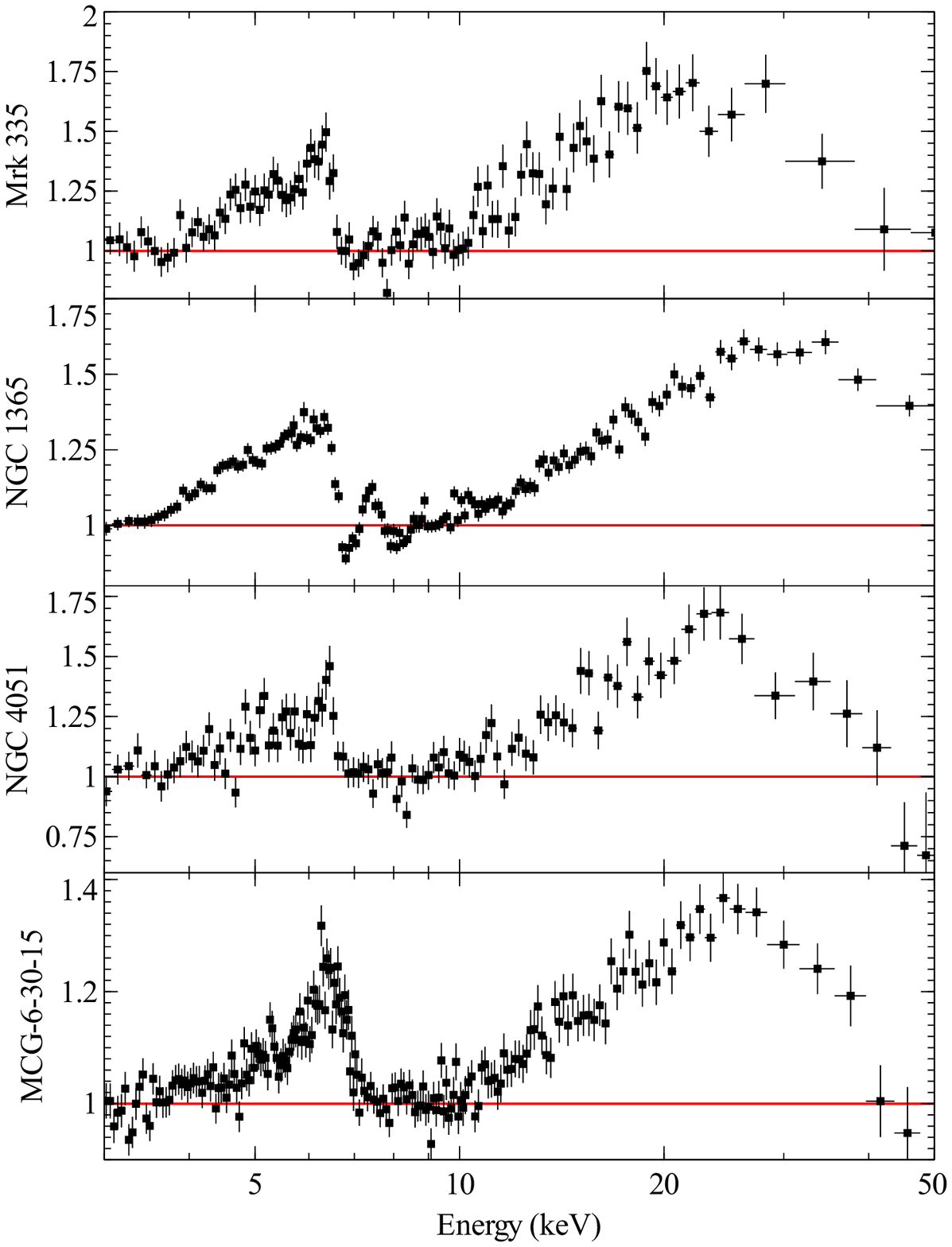}
\includegraphics[width=0.95\columnwidth]{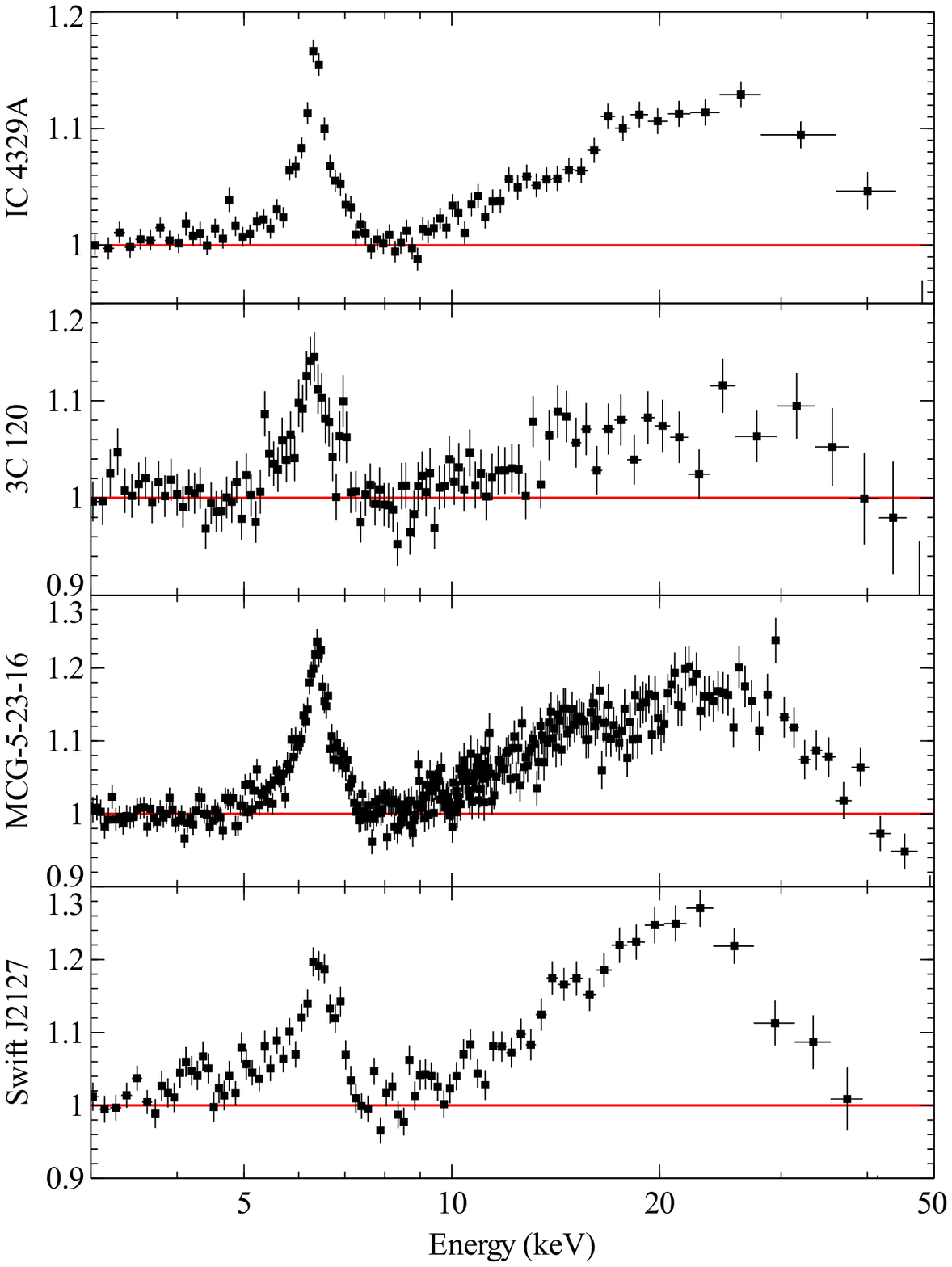}
\caption{NuSTAR spectral ratios to a simple power-law fitted between
  3--4 keV and 8--10 keV (from Fabian et al 2014). The broad iron line 
and Compton hump
  in the reflection spectrum are evident.
}
\label{label1}
\end{figure}

\begin{figure}
\includegraphics[width=0.98\columnwidth]{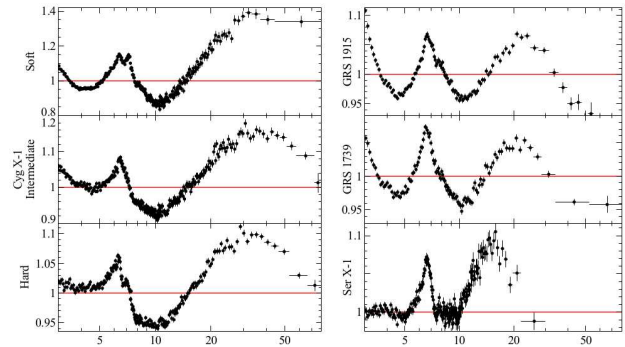}
\caption{NuSTAR spectral ratios for BHB (see e.g. Tomsick et al 2014;
  Miller et al 2013a, 2013b, 2015). 
}
\label{label1}
\end{figure}

\begin{figure}
\includegraphics[width=0.95\columnwidth]{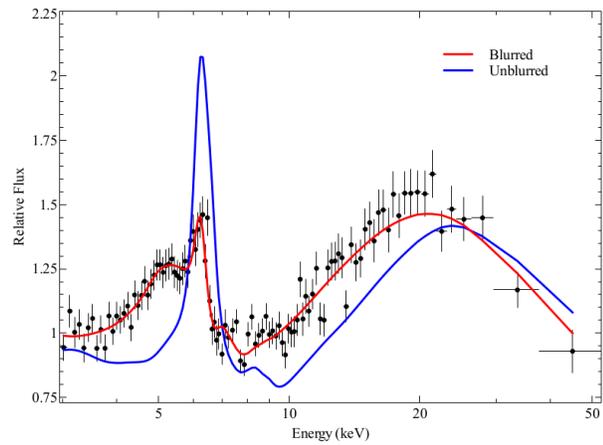}
\caption{Blurred and unblurred model spectra overplotted on NuSTAR
  data from the low state of Mkn335 (Parker et al 2014).
}
\label{label1}
\end{figure}

\begin{figure}
\includegraphics[width=0.95\columnwidth]{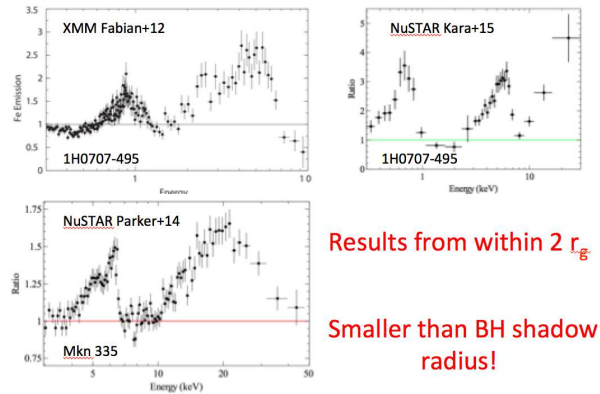}
\caption{Collection of AGN low state spectral ratios when the spectra
  are reflection dominated. Most of the emission from these spectra
  emerge from within $2r_{\rm g}$. 
}
\label{label1}
\end{figure}

A number of AGN occasionally drop into a reflection-dominated low
state\footnote{This is not the same phenomenon as the BHB low
  state}. Recent examples include 1H0707-495 (Fabian et al 2012; Kara
et al 2015) and Mkn 335 (Parker et al 2014; Figs~5 and 6). In these
cases the evidence indicates that the primary power-law is still
present and irradiating the disc strongly but appears weak in our
sightline due to a low coronal height which causes very strong light
bending. When the corona is close to the black hole the strong gravity
bends the power-law continuum away from our line of sight and down
onto the disc. Most of the reflection-dominated X-ray spectrum that we
observe originates from within a radius of 3 or even just
2$r_{\rm g}$. 

\section{X-ray Reverberation}

\begin{figure}
\includegraphics[width=0.95\columnwidth]{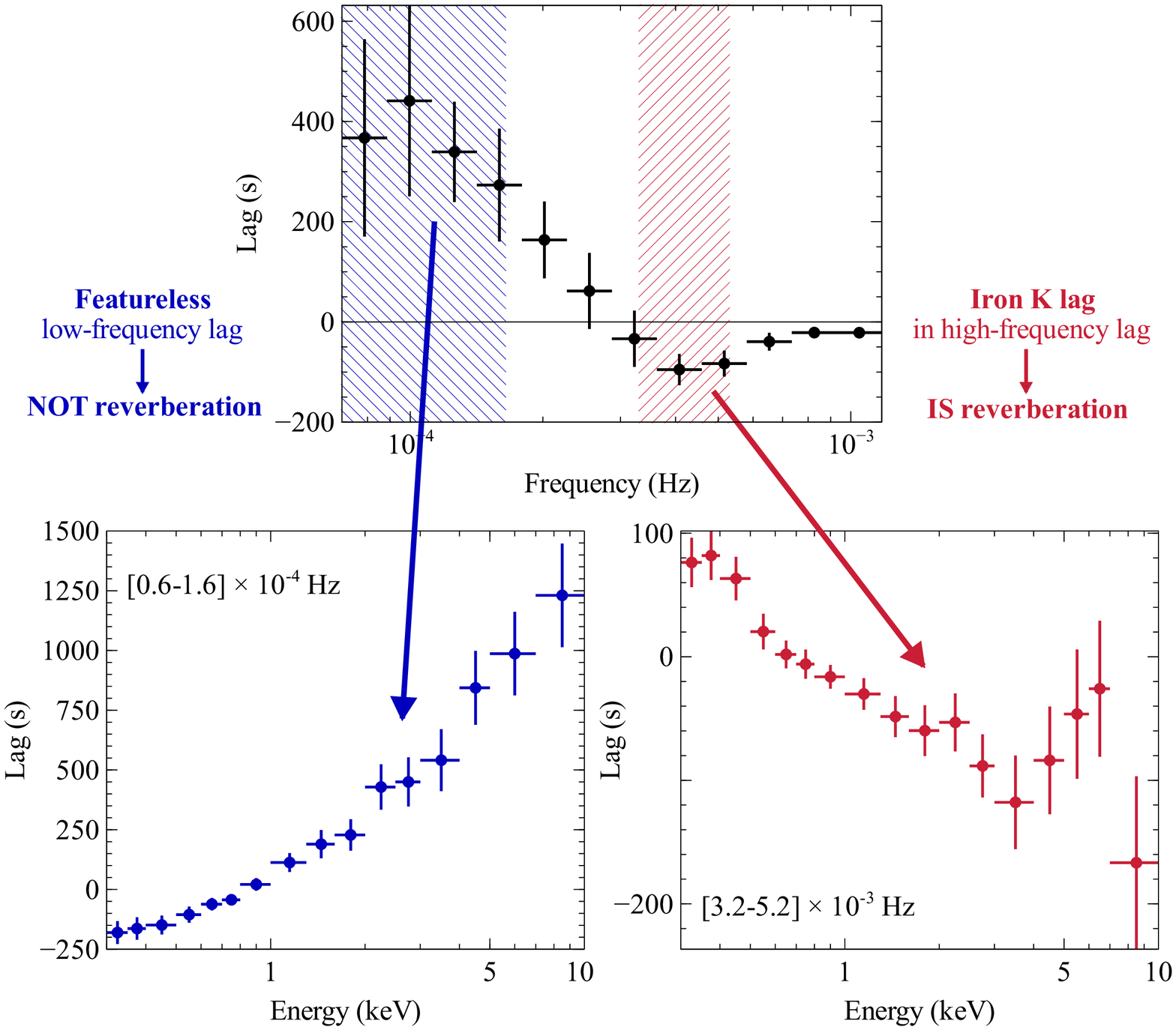}
\caption{Lag-frequency and lag-energy spectra of Akn 564 (Kara et
  al 2014). Fe K reflection features are seen in the high frequency
  lag-energy plot confirming that it is due to reverberation of the
  reflection spectrum.
}
\label{label1}
\end{figure}

\begin{figure}
\includegraphics[width=0.95\columnwidth]{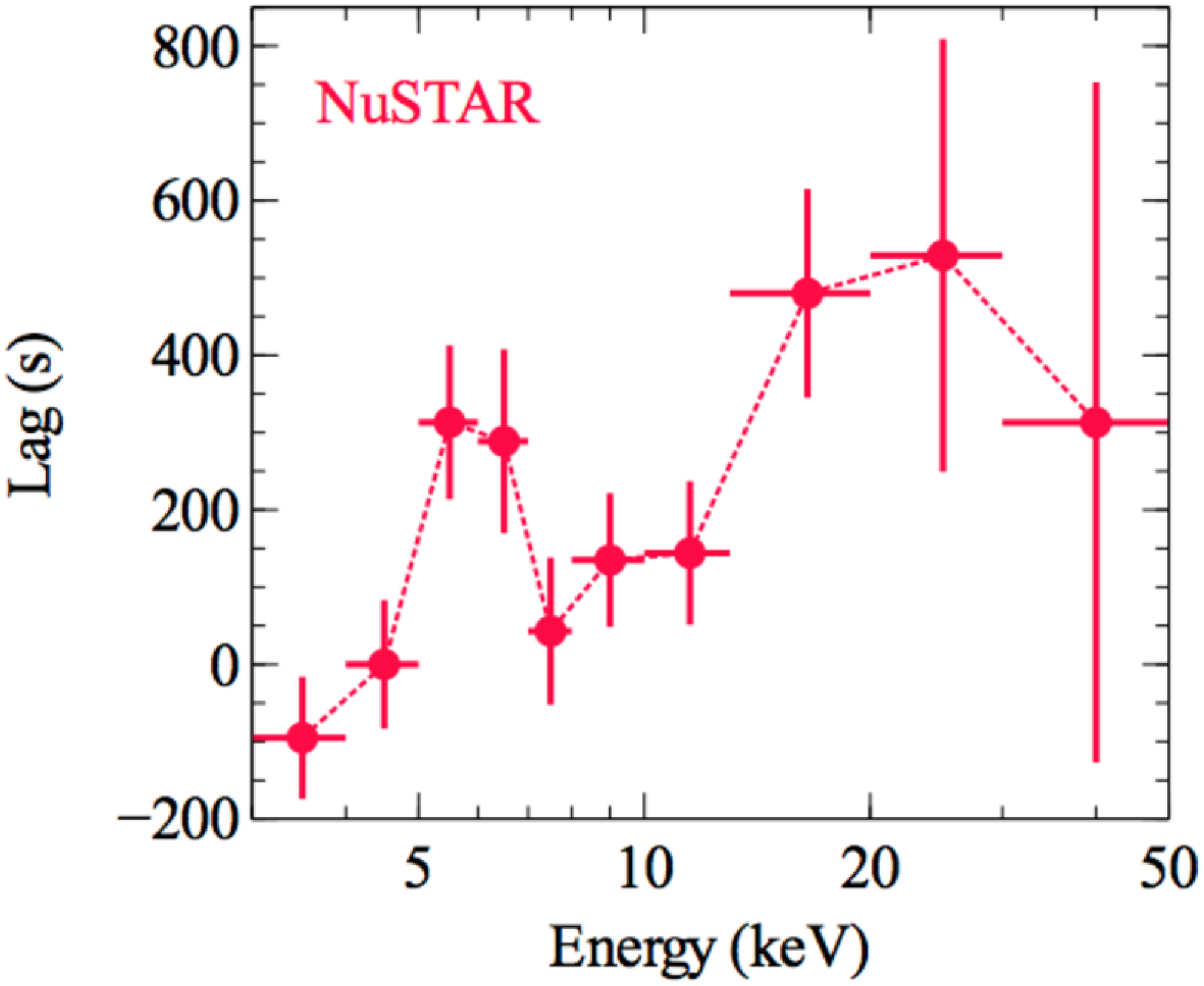}
\caption{Lag-energy plot of SWIFT J2127 (Kara et al 2015) showing both
  the broad iron line and Compton hump.
}
\label{label1}
\end{figure}

As already mentioned, there is a time lag between the detection of
variations in the primary coronal power-law and variation induced in
the reflection spectrum due to the greater path length taken by the
photons. Such reverberation provides us with a complementary mapping
device for understanding the inner regions of the accretion
flow (for a review see Uttley et al 2014).

Since their discovery using XMM data of 1H0707-495 (Fabian et al
2009), high-frequency lags attributable to reverberation are now
commonly seen in X-ray bright AGN which have strong rapid variability
(e.g. De Marco et al 2013). Lag energy spectra reveal clear evidence
that reverberation is involved through detection of the Fe K imprint
of the reflection spectrum (Fig.~7). Low frequency lags are also seen in many
sources (including BHB in which they were first detected back in the
late 1980s, Miyamoto \& Kitamoto 1989) but their spectrum tends to be
a rising power-law. They are probably due to small changes in the
corona, and thus the power-law continuum.

The energy lag spectrum of both Fe K and Compton hump has now been
obtained from NuSTAR, such as that shown in Fig.~7 by Kara et al
(2014) using code from Zoghbi et al (2013). Since we cannot separate
the direct primary continuum from the reflection spectrum, but observe
both mixed together, the observed lag spectrum is diluted by the
relative amount of reflection to power-law in each band. This must be
taken into account when modelling the lag spectrum. The observed lag
spectrum (e.g. Fig.~8) indicates that the response is seen first in
the red wing of both the iron line and Compton hump and later in the
less blurred parts which originate further out in the disc.  In
general, modelling (e.g. Cackett et al 2015; Emmanoulopoulos et al
2014) suggests that the corona is located at 3--10$r_{\rm ISCO}$.
  
No alternative model for broad iron lines is able to replicate the
full spectral timing results including the high frequency Fe K lags.

\section{Black Hole Spin}

\begin{figure}
\includegraphics[width=0.95\columnwidth]{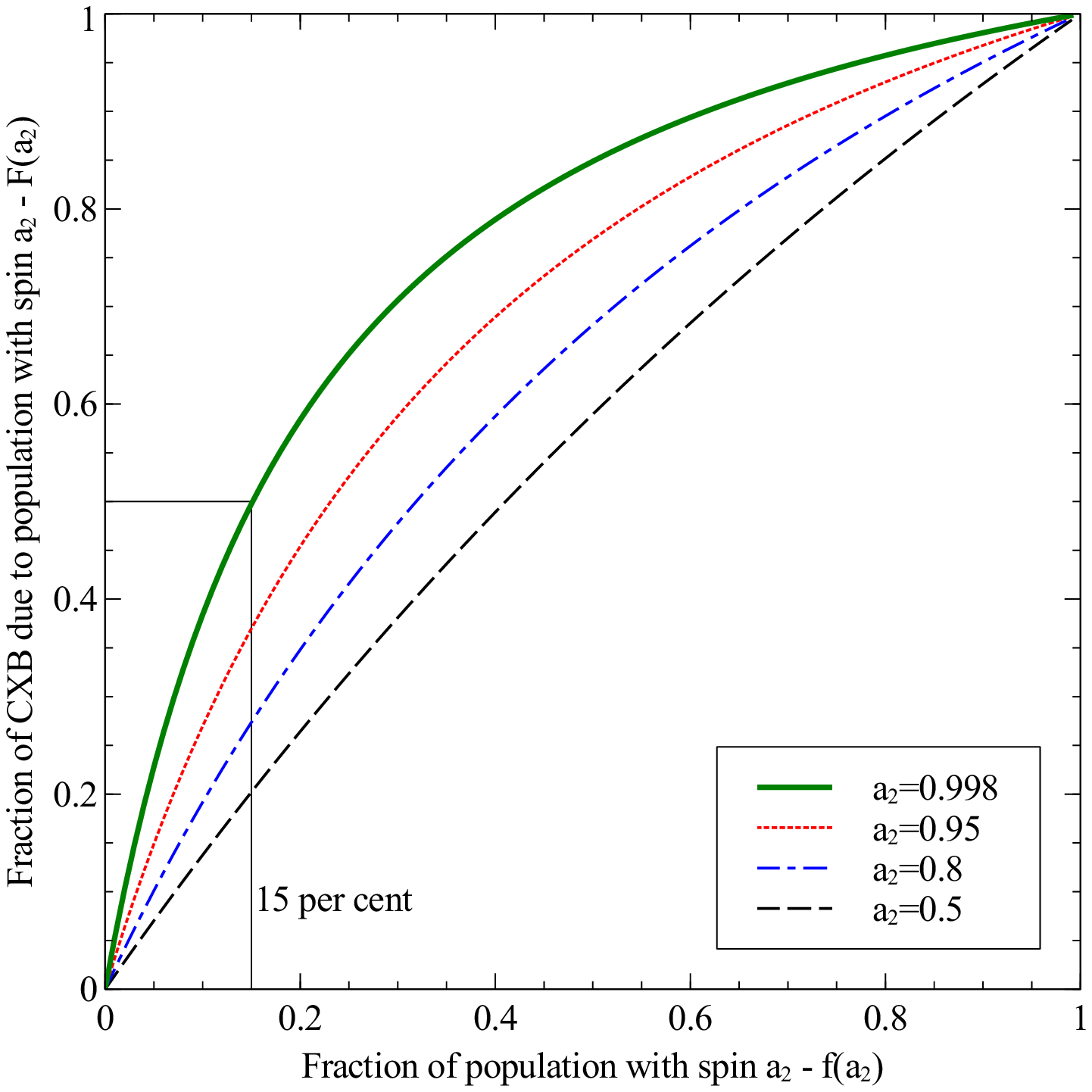}
\caption{Spin bias in the XRB intensity Vasudevan et al 2015). The
  plot assumes a two component AGN population, one fraction with spin
  zero the other with spin $a_2$. Only 15\% of the population is
  needed with maximal spin to give half the background.  }
\label{label1}
\end{figure}

\begin{figure}
\includegraphics[width=0.95\columnwidth]{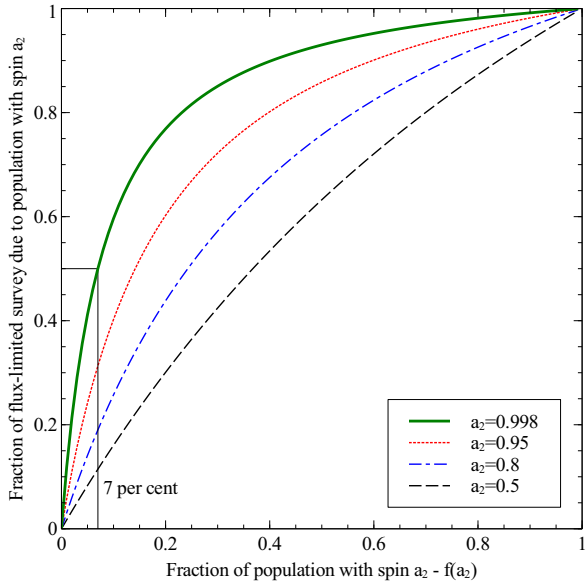}
\caption{Spin bias in source counts (Brenneman et al 2011; Vasudevan
  et al 2015). ONly 7\% ofb the population is needed with maximal
  spin to give half the bright sources counts.
}
\label{label1}
\end{figure}

The spin of the black hole can be determined if we identify the
innermost radius required by the blurring with the ISCO. Many of the
AGN sources from which the spin has been determined in this way have a
spin $> 0.9$, with some consistent with maximal spin of 0.998 (see
talk by Chris Reynolds and Reynolds 2014). This could be considered a
possible problem for the method if a wide range of spin is
expected. However, there is a strong bias towards detecting objects
with high spin due to the increase in radiative efficiency with spin,
which can be up to a factor of 5 (Brenneman et al 2011, Vasudevan et
al 2015).

To understand this bias, consider first the contribution of spinning
objects to the X-ray Background. Assume that all AGN are randomly
distributed between zero spin with accretion efficiency of 5.7\% and
near maximal spin with an efficiency about 5 times larger. Then if
just a random 1/6 of the AGN have high spin then together they produce
as much background intensity as the other 5/6 (Fig.~9).  This means
that half the background comes from 1/6 of the sources. We make no
assumptions here about luminosity, mass or distance, just that the
distribution of high spin among sources is completely random. The
30~keV spectral peak in the XRB may thus be due to the Compton hump in
reflection from a minority population of rapidly spinning black holes.

Carrying out a similar exercise for source counts is a little more
complicated and involves the 3/2 power. Only 1 in 14 sources, again
distributed randomly, need have high spin in order for them to produce
half the bright sources counts (Fig.~10; Vasudevan et al 2015).  The
two component model discussed here is obviously extreme. The point is
that if there is a broad intrinsic spin distribution then it should be
no surprise that the brightest AGN in the Sky host some of the most
rapidly spinning black holes.

\section{Coronal Physics}

Studies from several directions are revealing that corona are both
physically and radiatively compact. Their physical size can be deduced
from the emissivity profile of blurred reflection (typically of the
broad iron line, Wilkins \& Fabian 2012), from reverberation discussed
above and also from microlensing studies (e.g. Morgan et al 2012; see
talk by G. Chartas).  All such methods point to coronal sizes of about
$10 r_{\rm g}$ or less. This means that the photon density in the
corona is extremely high and thus that the probability an energetic
photon trying to escape the region will collide with another photon is
also high. The probability is determined by the source compactness
(Guilbert et al 1983)
\begin{equation}
\ell={L\over R}{\sigma_T\over {m_{\rm e}c^3}},
\end{equation}
where $L$ and $R$ are the luminosity and size of the region,
respectively.  If the product of the two photon energies exceeds
$2m_{\rm e} c^2$ an electron-positron pair can result. The particles
then help generate more photons and a runaway situation ensues which is
only stabilised if the conditions change to reduce the photon
energies.  The net result is a maximum temperature for the region
which drops lower as the compactness increases (e.g. Svensson 1984;
Stern et al 1995; Fig.~11). Recent observations now indicate that
$\ell\sim10-1000$, which is about an order of magnitude larger than
early indications based on source variability (Done \& Fabian 1989).

The temperature of the corona is determined from the high energy
cutoff of the power-law continuum. NuSTAR has made observations of the
cutoff fairly routine for bright AGN and Comptonization modelling
yielding coronal temperatures
$\Theta = kT/{m_{\rm e}c^2} \sim 0.03-0.8$.  Interestingly an inverse
correlation between $\ell$ and $\Theta$ is seen in the data, as
predicted by pair production models (Fig.~12; Fabian et al 2015). It is
therefore plausible that pair production controls the physics of the
corona. The corona itself may be a pair plasma and the upscattered
photons may originate from the magnetic field (i.e. as virtual
cycltron photons).

\begin{figure}
\includegraphics[width=0.95\columnwidth]{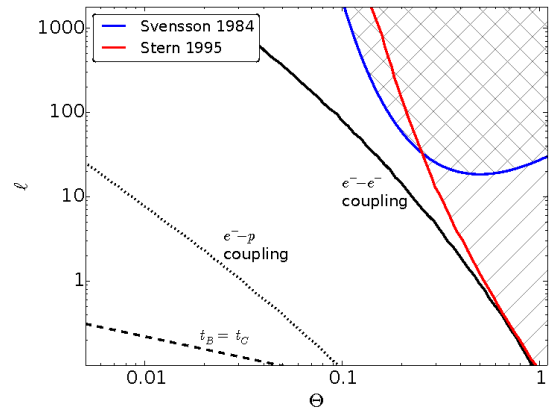}
\caption{$\ell - \Theta$ (compactness vs temperature) diagram showing
  key theoretical constraints. Compton cooling is faster than a)
  bremsstrahlung above the dashed line, b) electron--proton thermal
  coulping above the dotted line and c) electron--electron thermal
  coupling above the solid black line. Runaway pair production occurs 
above the blue and red lines according to the geometry assumptions made by
Svensson (1984) and Stern et al (1995). Sources are not expected to be
observed in the pair runaway region. 
}
\label{label1}
\end{figure}

\begin{figure}
\includegraphics[width=0.95\columnwidth]{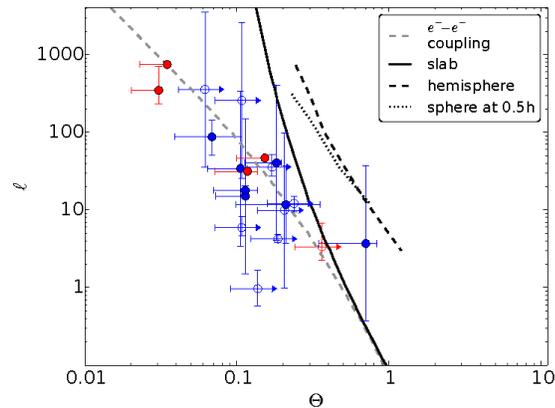}
\caption{$\ell - \Theta$ diagram for sources observed by NuSTAR (red
  for BHB and blue for AGN, Fabian et al 2015).  As expected, sources
  avoid the pair production region. When the temperature estimates are
  corrected for gravitational redshift the points shift closer to the
  pair production line.  Their proximity to the line suggests that
  pair production plays a key role in the behaviour of the corona.  }
\label{label1}
\end{figure}

\section{Discussion}

The innermost extremes of black hole accretion can be very extreme in some
currently observed AGN and BHB. Most of the X-ray luminosity
originates in a powerful compact corona lying along or close to the
spin axis of the black hole above (and below) the accretion disc. The
corona may not be static but rotate with the disc and move vertically
or expand horizontally. The particles within the corona may flow
upward (see for example the inset in Fig.~1 and models of Beloborodov
1999 and Malzac 2001).  X-ray spectral timing has enabled us to map
the innermost regions of the accretion flow and corona.

Current observations are photon-limited and provide only snapshots of
coronal behaviour in the X-ray brightest AGN. Understanding how the
corona is powered and operates is a major step in understanding how
the central engine works in the most luminous persistent objects in
the Universe.  More extensive long observations with XMM of the most
variable X-ray bright AGN will initiate the study of coronal behaviour
in a dynamical sense and demonstrate the coupling between disc and
corona. Low state AGN observations lack photons which again can be
compensated for with long observations. A very deep look at the low
state of an AGN with a rapidly spinning black hole (e.g. Fig.~5),
where most of the detected photons orginate in a reflection spectrum
from within $1r_{\rm g}$ of the horizon, will surely be worthwhile.

In the near future the broad band coverage of ASTRO-H, which is due
for launch in early 2016, will further reveal the corona and
reflection components of BHB and AGN. Its high resolution X-ray
microcalorimeter will enable clear separation of the effects of
intervening absorbing gas and outflows. Later in the 2020s, Athena
will make detailed soft X-ray timing spectroscopy a routine exercise
for many BHB and AGN.

\section{Acknowledgments}
Thanks to students, colleagues, collaborators and the NuSTAR team for
much work and many discussions on the topics covered here.  Michael
Parker is thanked for Figs 3 and 4. The European Research Commission
is thanked for Advanced Grant FEEDBACK.


\begin{thebibliography}{}
\bibitem{} Beloborodov, A.,: 1999, ApJ, 510, L133
\bibitem{} Brenneman L., et al., 2011, ApJ, 736, 103
\bibitem{} Cackett E.M., et al.: 2014, MNRAS, 438, 2980
\bibitem{} Dauser T., et al.,  2013, MNRAS, 430, 1694
\bibitem{} De Marco B., et al.,:2013, MNRAS, 431, 2441  
\bibitem{} Done, C., Fabian, A.C.,:1989, MNRAS, 240, 81
\bibitem{} Emmanoulopoulos, D., Papadakis, I.E., Dovciak, M., McHardy,
  I.M., 2014, MNRAS, 439, 3931 
\bibitem{} Fabian, A.C., Rees, M.J., Stella, L., White, N.E.,: 1989,
  MNRAS, 238, 729
\bibitem{} Fabian, A.C., et. al.: 2009, Nature 459, 540
\bibitem{} Fabian, A.C., et al.: 2012, MNRAS,419, 116 
\bibitem{} Fabian, A.C., Kara, E., Parker, M.L.,: 2014,
  Suzaku-MAXI2014, eds Ishida M., Petre R., Mitsuda K., p279, arXiv:1405.4150 
\bibitem{} Fabian, A.C., Lohfink, A., Kara, E., Parker, M.L.,
  Vasudevan, R., Reynolds, C.S., 2015, MNRAS, 451, 4375
\bibitem{} Garcia, J., Kallman, T.R., Muchotzky, R.F.,: 2011, ApJ,
  731, 131 
\bibitem{} Garcia, J., et al.: 2013, ApJ, 768, 146  
\bibitem{} Guilbert, P., Fabian, A.C., Rees, M.J.,: MNRAS, 205, 593
\bibitem{} Harrison, F., et al 2013, ApJ,770, 103 
\bibitem{} Kara, E., et al., 2013, MNRAS,434, 1129 
\bibitem{} Kara, E.,  et al., 2015, MNRAS, 446, 737 
\bibitem{} Kara, E., et al.,: 2015, MNRAS, 449, 234
\bibitem{} Lightman, A.P., White, T,R.,: 1988, ApJ, 335, 57
\bibitem{} Malzac, J., Beloborodov, A., Poutanen, J.,: 2001,
  MNRAS, 326, 417
\bibitem{} Marinucci, A., et al., 2014, MNRAS, 440, 2347
\bibitem{} Miller, J.M., et al., 2013a, ApJL, 775, L45
\bibitem{} Miller, J.M., et al., 2013b, ApJL, 779, L2
\bibitem{} Miller, J.M., et al., 2013a, ApJL, 799, L6
\bibitem{} Miyamoto, S., Kitamoto, S.,: 1989, Nature, 342, 773
\bibitem{} Morgan C.W., et al.: 2012, ApJ,756, 52 
\bibitem{} Parker, M.L.,et al.: 2014, MNRAS, 443, 1723
\bibitem{} Risaliti, G., et al.: 2013, Nature, 494, 449
\bibitem{} Ross, R.R., Fabian, A.C.,: 1993, MNRAS, 261, 74
\bibitem{} Ross, R.R., Fabian, A.C.,: 2005, MNRAS, 358, 211
\bibitem{} Stern, B., Poutanen, J., Svensson, R., Sikora M., Begelman,
  M.C.,: 1995, ApJ, 449 
\bibitem{} Svensson, R.,: 1984, MNRAS, 209, 1745
\bibitem{} Thorne, K.,: 1974, ApJ, 191, 507 
\bibitem{} Tomsick, J., et al.,: 2014, ApJ, 780, 78
\bibitem{} Uttley, P., Cackett, E.M., Fabian, A.C., Kara, E., Wilkins,
  D.R.,: 2014, A\&ARv, 22, 72 
\bibitem{} Vasudevan, R., Fabian, A.C., Reynolds, C.S., Dauser, T.,
  Gallo, L.,: 2015, arXiv:1506.01027
\bibitem{} Walton, D.J., et al.,: 2014, ApJ, 788, 76
\bibitem{} Wilkins, D.R., Fabian A.C.,: MNRAS, 414, 1269
\bibitem{} Zoghbi A., Reynolds, C., Cackett, E.M.,: 2013, ApJ, 777, 24 

\end{thebibliography}
\end{document}